\documentclass{andp2012}
\usepackage[english]{babel}

\usepackage{amsmath,graphicx,epsfig}
\usepackage{epstopdf}
\usepackage{euscript}
\usepackage{amsfonts}
\usepackage{amssymb}
\usepackage{float}

\def\abrk#1{{\langle#1\rangle}}

\def\bra#1{{\langle#1|}}

\def\cg(#1,#2)(#3,#4)(#5,#6){\bra{#1,#2,#3,#4}#5,#6\rangle}

\def\threej(#1,#2)(#3,#4)(#5,#6){\begin{pmatrix}#1&#3&#5\\#2&#4&#6\end{pmatrix}}
\def\sixj(#1,#2,#3)(#4,#5,#6){\begin{Bmatrix}#1&#2&#3\\#4&#5&#6\end{Bmatrix}}
\def\ninej(#1,#2,#3)(#4,#5,#6)(#7,#8,#9){\begin{Bmatrix}#1&#2&#3\\#4&#5&#6\\#7&#8&#9\end{Bmatrix}}

\def\sR{{\ensuremath{\EuScript R}}}

\def\mb{\mathbf}
\def\bs{\boldsymbol}
\def\mc{\mathcal}

\keywords{Magnetic moments, rubidium, gyromagnetic ratios, magnetometry.}
\title{Measurement of the ratio between g-factors of the ground states of $^{87}$Rb and $^{85}$Rb}
\author[J. Mora]{Jason Mora\inst{1}}
\author[A. Cobos]{Aracely Cobos\inst{1}}
\author[D. Fuentes]{Dominic Fuentes\inst{1}}
\author[D. F. Jackson Kimball]{Derek F. Jackson Kimball\inst{1,}\footnote{Corresponding author\quad E-mail:~\textsf{derek.jacksonkimball@csueastbay.edu}}}
\address[1]{Department of Physics, California State University -- East Bay, Hayward, California 94542-3084, USA}
\shortauthors{J. Mora et al.}
\begin{abstract}
The ratio between the Land\'e g-factors of the $^{87}$Rb $F=2$ and $^{85}$Rb $F=3$ ground state hyperfine levels is experimentally measured to be $g_F(87)/g_F(85) = 1.4988586(1)$, consistent with previous measurements. The g-factor ratio is determined by comparing the Larmor frequencies of overlapping ensembles of $^{87}$Rb and $^{85}$Rb atoms contained within an evacuated, antirelaxation-coated vapor cell. The atomic spins are polarized via synchronous optical pumping and the Larmor frequencies are measured by off-resonant probing using optical rotation of linearly polarized light. The accuracy of this measurement of $g_F(87)/g_F(85)$ exceeds that of previous measurements by a factor of $\approx 50$ and is sensitive to effects related to quantum electrodynamics.
\end{abstract}
\shortabstract
\begin{document}
\maketitle

\section{Introduction}
\label{Sec:Introduction}

Precision measurements of atomic structure offer insights into fundamental interactions within atomic systems and can be used to search for new physics \cite{Saf18-review}. In particular, comparisons between measurements and calculations of atomic structure inform the development of atomic theory \cite{karshenboim2005precision,roberts2015parity} and can constrain exotic interactions between atomic constituents \cite{Fic17,ficek2018constraints}. Theoretical calculations for alkali atoms are particularly advanced due to their relatively simple atomic structure and are motivated by their widespread application in clocks and frequency standards \cite{santarelli1999quantum,lombardi2007nist}, atom cooling and trapping \cite{wieman1999atom}, experiments with quantum degenerate gases \cite{anderson1995observation,demarco1999onset}, atomic magnetometry \cite{Bud02,Bud13}, measurements of parity-violation \cite{wood1997measurement,guena2005measurement}, and searches for permanent electric dipole moments (EDMs) \cite{murthy1989new,nataraj2008intrinsic}.

Recently, we carried out a search for a coupling between rubidium (Rb) spins and the gravitational field of the Earth \cite{Kim17-GDM} by simultaneously measuring the spin precession frequencies of $^{87}$Rb atoms in the $F=2$ ground-state hyperfine level and $^{85}$Rb atoms in the $F=3$ ground-state hyperfine level, denoted $\Omega_{87}$ and $\Omega_{85}$, respectively. Since the precession is due predominantly to the Zeeman interaction of the Rb spins with an applied magnetic field of magnitude $B$, the precession frequency for a given isotope $i$ is\footnote{The measured spin precession frequency $\Omega_i$ is only approximately equal to the Larmor frequency due to both higher-order-terms and non-magnetic effects that can also lead to spin precession as discussed in Ref.~\cite{Kim17-GDM}. However, as discussed here and also in Ref.~\cite{Kim17-GDM}, in our experiment these effects are controlled at a level much better than a part-per-million.}
\begin{align}
\Omega_i \approx \frac{g_F(i) \mu_B B}{\hbar}~,
\label{Eq:Larmor-freq}
\end{align}
where $g_F(i)$ is the Land\'{e} factor for isotope $i$ in the ground-state hyperfine level with total atomic angular momentum $F$, $\mu_B$ is the Bohr magneton, and $\hbar$ is Planck's constant. Therefore the data acquired in the experiment described in Ref.~\cite{Kim17-GDM} can naturally be used to determine the ratio $\mc{R}$ between the ground-state Land\'{e} g-factors of the two Rb isotopes:
\begin{align}
\mc{R} = \frac{\Omega_{87}}{\Omega_{85}} \approx \frac{g_F(87)}{g_F(85)}~.
\label{Eq:g-factor-ratio}
\end{align}
In this work, we carry out analysis of the data from Ref.~\cite{Kim17-GDM} in order to determine $\mc{R}$ with an accuracy exceeding that of past measurements \cite{Whi68-Rb-g-factor-ratio-measurement} by a factor of $\approx 50$. The experimentally determined value for $\mc{R}$ can be compared to atomic calculations, and is of sufficient accuracy to be sensitive to quantum electrodynamic (QED) corrections to the Hamiltonian describing magnetic interactions \cite{Ant94-Relativistic-corrections-to-g-factors,Chan11-cold-Rb-g-factor-expt}. Accurate understanding of Rb g-factors may be useful, for example, in magnetic field measurements where the field is determined by observing Zeeman shifts \cite{ciampini2017optical,george2017pulsed} or spin-precession frequencies \cite{Bud02,Bud13}.

\section{Theory}
\label{Sec:Theory}

The Land\'{e} g-factor $g_F$ describes the relationship between the angular momentum $\mb{F}$ and the magnetic moment $\bs{\mu}$ of a particle,
\begin{align}
\bs{\mu} = -g_F \mu_B \mb{F}~.
\end{align}
Due to the $\bs{\mu}\times\mb{B}$ torque generated by a magnetic field $\mb{B}$, particles precess at the Larmor frequency $\Omega$ given in Eq.~\eqref{Eq:Larmor-freq}. For a composite particle such as an atom, there are multiple contributions to the total angular momentum and magnetic moment from the electronic spin $\mb{S}$ and orbital angular momentum $\mb{L}$ as well as the nuclear spin $\mb{I}$. In the present experiment, the $^2$S$_{1/2}$ ground states of $^{87}$Rb and $^{85}$Rb are studied where $L=0$, and thus to lowest order,
\begin{align}
\bs{\mu} & = -g_S \mu_B \mb{S} + g_I \mu_B \mb{I} \\
& = -\frac{g_S\abrk{ \mb{S}\cdot\mb{F} } - g_I\abrk{ \mb{I}\cdot\mb{F} } }{F(F+1)} \mu_B\mb{F}~,
\label{Eq:magnetic-moment}
\end{align}
where $g_S$ is the electronic g-factor, $g_I$ is the nuclear g-factor, and $\abrk{\cdots}$ denotes the expectation value of the considered quantity. Thus the atomic g-factor $g_F$ is given by
\begin{align}
g_F =& g_S \frac{F(F+1)+S(S+1)-I(I+1)}{2F(F+1)} \nonumber \\
&- g_I \frac{F(F+1)+S(S+1)-I(I+1)}{2F(F+1)}~.
\label{Eq:gF}
\end{align}

Although the g-factor for a free electron, $$g_e = 2.00231930436146(56)~,$$ is one of the most accurately known physical constants \cite{Han08-electron-g-factor}, $g_S$ for an electron in a bound state of an atomic system is modified from $g_e$ at about a few parts-per-million level by various relativistic and QED corrections to the Hamiltonian describing magnetic interactions \cite{Ant94-Relativistic-corrections-to-g-factors,Chan11-cold-Rb-g-factor-expt}. Thus a measurement of Rb g-factor ratios with accuracy significantly below the parts per million level is sensitive to such relativistic and QED corrections.

As noted in the introduction, we have recently simultaneously measured spin precession frequencies of the $^{87}$Rb $F=2$ and $^{85}$Rb $F=3$ ground state hyperfine levels for gas-phase ensembles of Rb atoms co-located in antirelaxation-coated vapor cells \cite{Kim17-GDM}. To the extent that non-magnetic causes of precession are minimized, the ratio of the precession frequencies $\mc{R}$ is a measure of the ratio of the atomic g-factors [Eq.~\eqref{Eq:g-factor-ratio}].

The best existing measurement of Rb g-factor ratios was reported in Ref.~\cite{Whi68-Rb-g-factor-ratio-measurement}, where Rb spin-precession was measured in ambient magnetic fields of $B \approx 50~{\rm G}$. The set of measurements described in Ref.~\cite{Whi68-Rb-g-factor-ratio-measurement} determined various ratios of g-factors for Rb electrons and nuclei, from which values of $g_S$ and $g_I$ can both be extracted at about the part-per-million level, yielding $\mc{R} = 1.498862(5)$ based on Eq.~\eqref{Eq:gF}, where the uncertainty in $\mc{R}$ is dominated by uncertainty in the $g_S$ values.

\section{Experimental Setup and Procedure}
\label{Sec:ExptSetup}

\begin{figure}
\includegraphics[width=3.7 in]{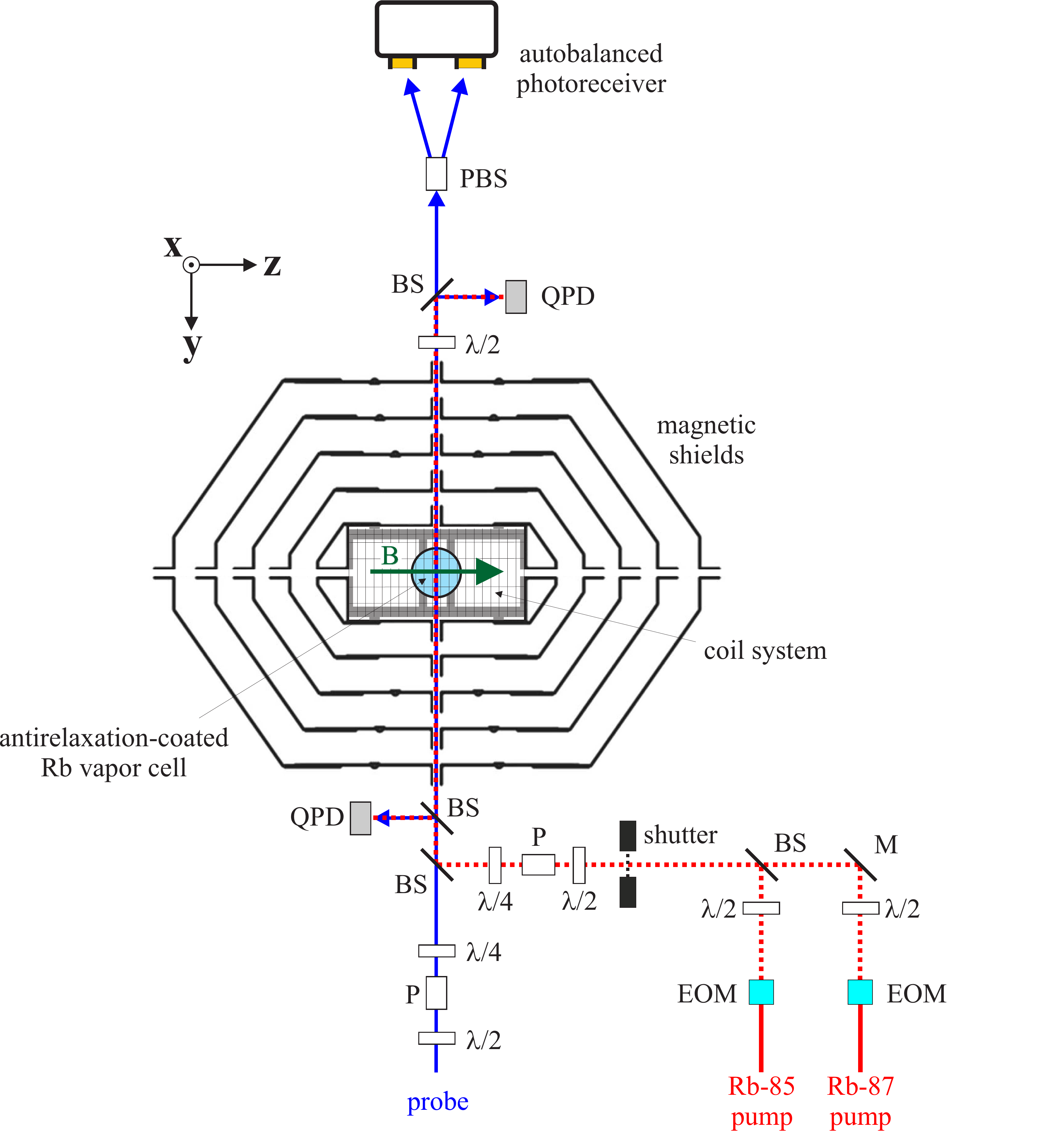}
\caption{Schematic of the experimental setup from Ref.~\cite{Kim17-GDM}. P~=~linear polarizer, M~=~mirror, BS~=~(nonpolarizing) beamsplitter, PBS~=~polarizing beamsplitter, $\lambda/4$~=~quarter-wave plate, $\lambda/2$~=~half-wave plate, EOM~=~electro-optic modulator, QPD~=~quadrant photodiode. Designation of $\mb{x}$, $\mb{y}$, and $\mb{z}$ directions is shown in the upper left corner. Red solid and dashed lines represent the pump beams, blue arrows represent the probe beam. The green arrow at the center of the diagram represents the applied magnetic field $\mb{B}$. Assorted optics and electronics for laser control, data acquisition, and experiment control are not pictured.}
\label{Fig:ExperimentalSetup}
\end{figure}

The experimental setup used to measure the spin precession frequencies of the $^{87}$Rb $F=2$ and $^{85}$Rb $F=3$ ground state hyperfine levels is described in detail in Refs.~\cite{Kim17-GDM,Kim13-GDMsetup}. Synchronous laser optical pumping generates precessing spin polarization of Rb atoms transverse to an applied magnetic field $\mb{B}$, and off-resonant laser light is used to simultaneously measure the spin precession frequencies of $^{85}$Rb and $^{87}$Rb (Fig.~\ref{Fig:ExperimentalSetup}). The vapor of Rb atoms is contained within a buffer-gas-free, antirelaxation-coated, spherical glass cell that is 5~cm in diameter. Both alkene-coated \cite{balabas2010polarized} and alkane-coated \cite{alexandrov2002light} cells were used to check for cell- and coating-related systematic errors during the experiments. The cell is located inside a set of nine independent magnetic field coils that enable control of longitudinal and transverse components of $\mb{B}$ as well as all first-order gradients and the second-order gradient along $\mb{B}$. The cell and coil system are nested within a temperature-stabilized, five-layer mu-metal shield system that provides near uniform shielding of external fields to a part in $10^7$ \cite{xu2006magnetic,Kim16}.

Measurement of $\Omega_{85}$ and $\Omega_{87}$ is carried out using a pump/probe sequence. During the 1~s duration pump stage, two collinear, circularly polarized laser beams propagate through the Rb vapor transverse to $\mb{B}$: one tuned to the center of the Doppler-broadened $^{85}$Rb $F=2 \rightarrow F'$ hyperfine component of the D2 transition and the other tuned to the center of the $^{87}$Rb $F=1 \rightarrow F'=2$ hyperfine component of the D1 transition. These laser beams optically pump atoms into the hyperfine levels of interest (which yield the largest optical rotation signals). The pump beams are independently amplitude-modulated by electro-optic modulators at frequencies matching the corresponding Larmor frequencies for the $^{85}$Rb $F=3$ and $^{87}$Rb $F=2$ ground state hyperfine levels, respectively. During the 1~s duration probe stage, the pump beams are shuttered, and optical rotation of a linearly polarized probe beam is measured with a polarizing beamsplitter and autobalanced photoreceiver. The probe beam is detuned several GHz to the low frequency side of the $^{87}$Rb D2 $F=2 \rightarrow F'$ transition and frequency stabilized using a wavemeter. At this detuning, spin precession of atoms in the $^{85}$Rb $F=3$ and $^{87}$Rb $F=2$ ground state hyperfine levels can be simultaneously measured.

A number of phenomena other than the linear Zeeman shift can affect the measured spin precession frequencies: these include magnetic field gradients \cite{cates1988relaxation,sheng2014new}, light shifts \cite{Mat68,Bul71,Coh72,Hap72}, asynchronous optical pumping \cite{Swa13}, spin-exchange collisions \cite{Hap72,Sch89,Dmi97,Dmi07}, the nonlinear Zeeman effect \cite{Aco06,Jen09,Cha10}, and the gyro-compass effect \cite{Ven92,Hec08,Bro10,Tul13}. A variety of experimental procedures and auxiliary measurements, described in detail in Ref.~\cite{Kim17-GDM}, were undertaken to cancel and control these effects. To minimize the magnetic field gradients that arise, for example, due to residual magnetization of the innermost shield layer, the widths of the magnetic resonance signals (which depend on the gradients) were minimized as a function of applied gradients -- both at zero field and at the nonzero applied fields at which the g-factor ratio measurements were performed. This procedure was estimated to reduce gradients to $\lesssim 5~{\rm \mu G/cm}$ in all directions, which translates to a systematic error in $\mc{R}$ of $\lesssim 2 \times 10^{-8}$ according to the analysis described in Ref.~\cite{cates1988relaxation}. Vector light shifts from the probe beam were zeroed using ellipticity-induced nonlinear magneto-optical resonances as described in Ref.~\cite{kimball2017situ}. Frequency shifts due to asynchronous optical pumping, which arise due to a magnetic resonance effect occurring when the pump modulation frequency is detuned from the Larmor frequency, are minimized by keeping the pump modulation frequencies within 3~mHz of the Larmor frequency with a feedback control loop. Since the $^{87}$Rb and $^{85}$Rb spins are precessing at different frequencies, spin-exchange frequency shifts from the predominantly transversely polarized (perpendicular to $\mb{B}$) Rb sample are negligible compared to other sources of error. However, it was discovered that non-negligible longitudinal spin-polarization (along $\mb{B}$) was produced by scattered pump light. Both the circular polarization of the pump light and the direction of $\mb{B}$ were reversed in the experiment which, in principle, averages out the systematic error in $\mc{R}$ due to the scattered pump light. However, the reversals of pump light polarization and $\mb{B}$ were not perfect, leading to an estimated systematic uncertainty in $\mc{R}$ of $\lesssim 4 \times 10^{-8}$. The nonlinear Zeeman effect was cancelled by choosing the power, polarization, and detuning of the probe beam such that the tensor light shifts compensated the nonlinear Zeeman effect \cite{Kim17-GDM}. Table~\ref{Table:systematic-errors} shows estimates of the dominant sources of systematic errors in determination of $\mc{R}$. When these systematic uncertainties are combined in quadrature, the overall systematic uncertainty in $\mc{R}$ is found to be $\approx 4.5 \times 10^{-8}$.

\begin{table}[h]
\centering
\begin{andptabular}[\linewidth]{lc}{Estimated upper limits on the most important contributions to systematic error in the determination of $\mc{R}$, see Ref.~\cite{Kim17-GDM} for details.}
\label{Table:systematic-errors}
Description & Effect on $\mc{R}$ ($\times 10^{-9}$)\\
\rule{0ex}{3.6ex} Scattered pump light along $\mb{B}$ & $40$  \\
\rule{0ex}{3.6ex} Magnetic field gradients & $20$ \\
\rule{0ex}{3.6ex} Asynchronous optical pumping & $8$  \\
\rule{0ex}{3.6ex} Tensor shifts & $2$ \\
\rule{0ex}{3.6ex} Cumulative systematic error & 45 \\
\end{andptabular}
\end{table}

The experimental procedures used to compensate and control these various systematic effects demanded that the chosen values of many of the experimental parameters were interconnected, which limited the ability to independently vary parameters. As a further check for unknown systematic effects, data were taken at two different magnetic field magnitudes: $B = 19.052729(3)~{\rm mG}$ (denoted the low field) and $B = 28.579094(3)~{\rm mG}$ (denoted the high field), where the field magnitudes were determined by measurement of $\Omega_{87}$. The probe beam power and detuning were adjusted accordingly for each field magnitude to compensate the tensor shifts. The data were collected in a series of experimental runs that consisted of a number of calibration measurements followed by 1280 individual acquisitions. Half of the acquisitions were taken with right-hand-circularly (RHC) polarized pump light and the other half with left-hand-circularly (LHC) polarized pump light (switching automatically every 40 acquisitions). Additionally, half of the acquisitions were taken with $\mb{B}$ pointing in the $+\mb{z}$ direction (toward the North star, along Earth's rotation axis) and half were taken with $\mb{B}$ pointing in the $-\mb{z}$ direction. The magnetic field was switched between the + and - directions once per run, and the magnetic field gradients were measured and compensated between every field switch. Experimental Runs 1-4, 8 and 9 were taken at the low field magnitude, while experimental runs 5-7 were taken at the high field magnitude. Although systematic errors related to wall collisions were estimated to be entirely negligible for our experiment in Ref.~\cite{Kim13-GDMsetup}, as a precaution data were taken using two different cells with different coatings: alkene \cite{balabas2010polarized} for Runs 1-7 and alkane \cite{alexandrov2002light} for Runs 8 and 9. The stems of the cells were also oriented differently between experimental runs so as to change the quadrupolar shape anisotropy \cite{Pec16}. As discussed in the next section, no statistically significant shifts of $\mc{R}$ were found from run-to-run.

\section{Analysis and Results}
\label{Sec:Results-Analysis}

\begin{figure}
\includegraphics[width=3.5 in]{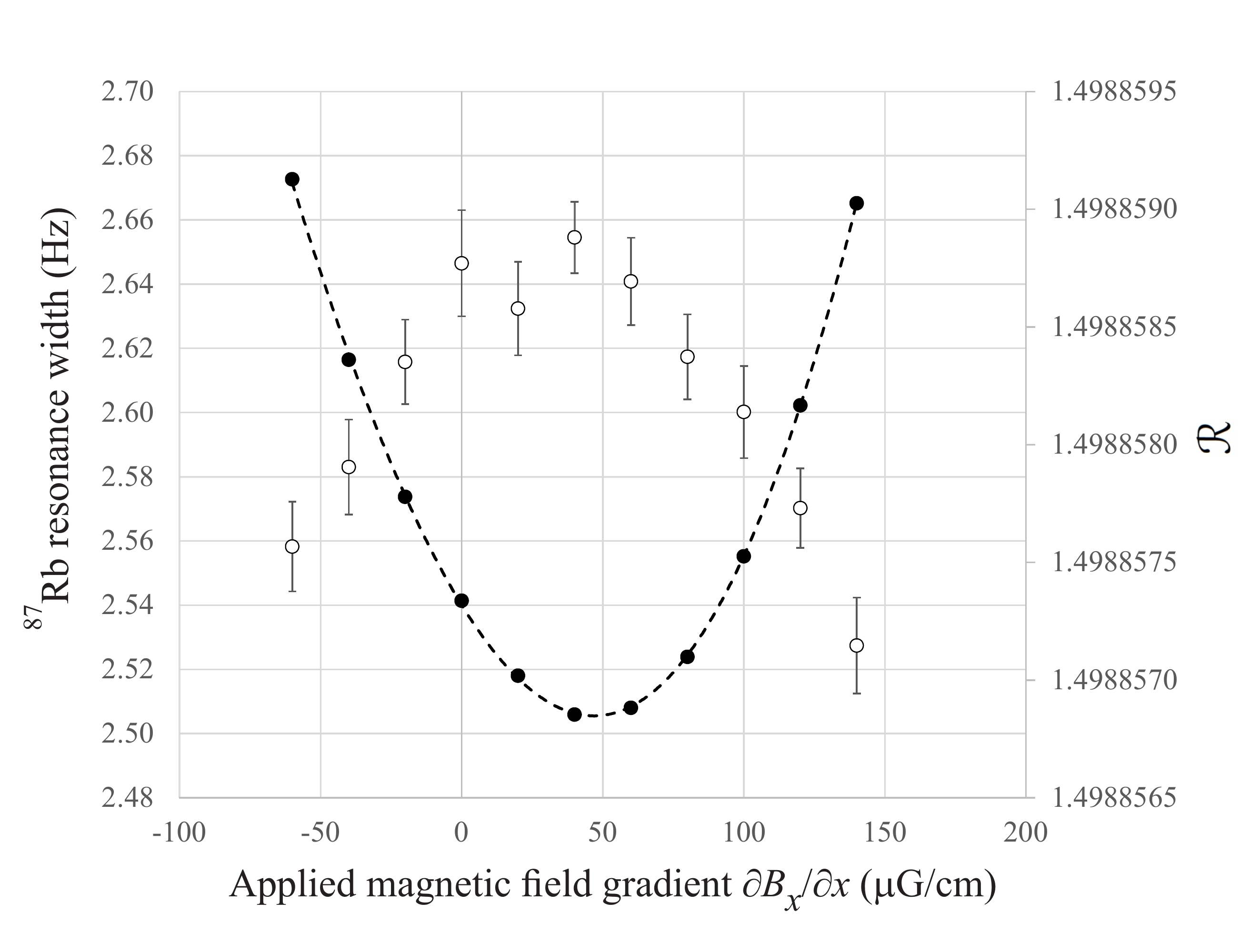}
\caption{Dependence of $\mc{R}$ (open circles) and the width of the Lorentzian fit to the $\Omega_{87}$ peak (filled circles) on applied first-order magnetic field gradient ($\partial B_x/\partial x$), where $\hat{\mb{x}}$ is orthogonal to the direction of $\mb{B}$ (whose direction is specified to be along $\pm\hat{\mb{z}}$) and the laser beam propagation direction (specified to be along $-\hat{\mb{y}}$), see Fig.~\ref{Fig:ExperimentalSetup}. For this data, the alkene-coated vapor cell was used and $B = 19.052729(3)~{\rm mG}$. Because the fractional effect of gradients on the width is larger than the fractional effect on $\mc{R}$, the gradients can be efficiently compensated by minimizing the widths. However, this experimental procedure does not eliminate the effect of stochastic glitches that shift the magnetic field gradient values.}
\label{Fig:GradientCompensationExample}
\end{figure}

A serious issue affecting the quality of the data was sporadic low-level power supply glitches causing temporary shifts of the currents through the coils controlling the leading magnetic field and magnetic field gradients. The noise was observed to be significantly larger when the power supply was set to generate a leading field in the $-\mb{z}$ direction and for larger currents. Also, there was no evidence of such noise in Runs 1 and 2, suggesting that perhaps this excess noise resulted from the failure of some component within the current supply, possibly induced by a series of power outages in the laboratory between Runs 2 and 3.  Although these sporadic glitches were rather small compared to the average applied current through the coils, at the level of the precision of our measurement it caused statistically significant shifts of precession frequencies. These precession frequency shifts arise due to gradient-induced geometric phase effects as described in Refs.~\cite{cates1988relaxation,sheng2014new} that do not exactly cancel in $\mc{R}$ (in spite of motional averaging in the evacuated cell \cite{Pus06b}). Worse yet, because the g-factor ratio depends on the square of the gradients as seen in Fig.~\ref{Fig:GradientCompensationExample}, the power supply glitches caused a systematic bias in the data sample because if the gradients randomly increased or decreased by a small amount as a result of current glitches, in either case the ratio $\mc{R}$ was shifted to smaller values. Thus random changes in the gradients do not average out in the determination of $\mc{R}$.

In the work described in Ref.~\cite{Kim17-GDM}, this issue was addressed by executing a strict cut on the data based on the quality of the fits as expressed through the uncertainty of the individual measurements of $\Omega_{87}$ and $\Omega_{85}$, which were found to be correlated with the glitches. In this reanalysis of the data from Ref.~\cite{Kim17-GDM}, we chose to approach the analysis in a different way.

\begin{figure}
\includegraphics[width=3.5 in]{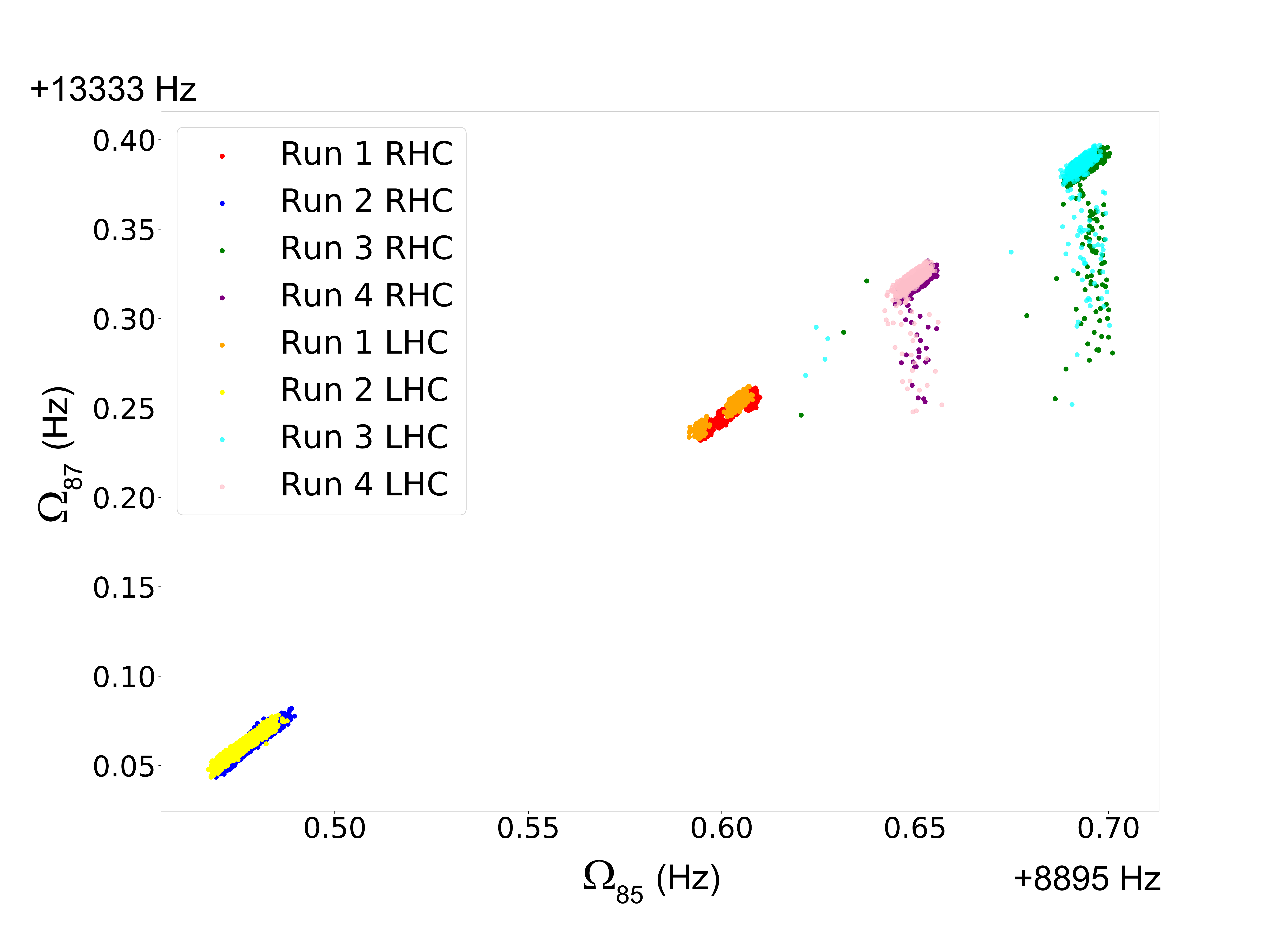}
\includegraphics[width=3.5 in]{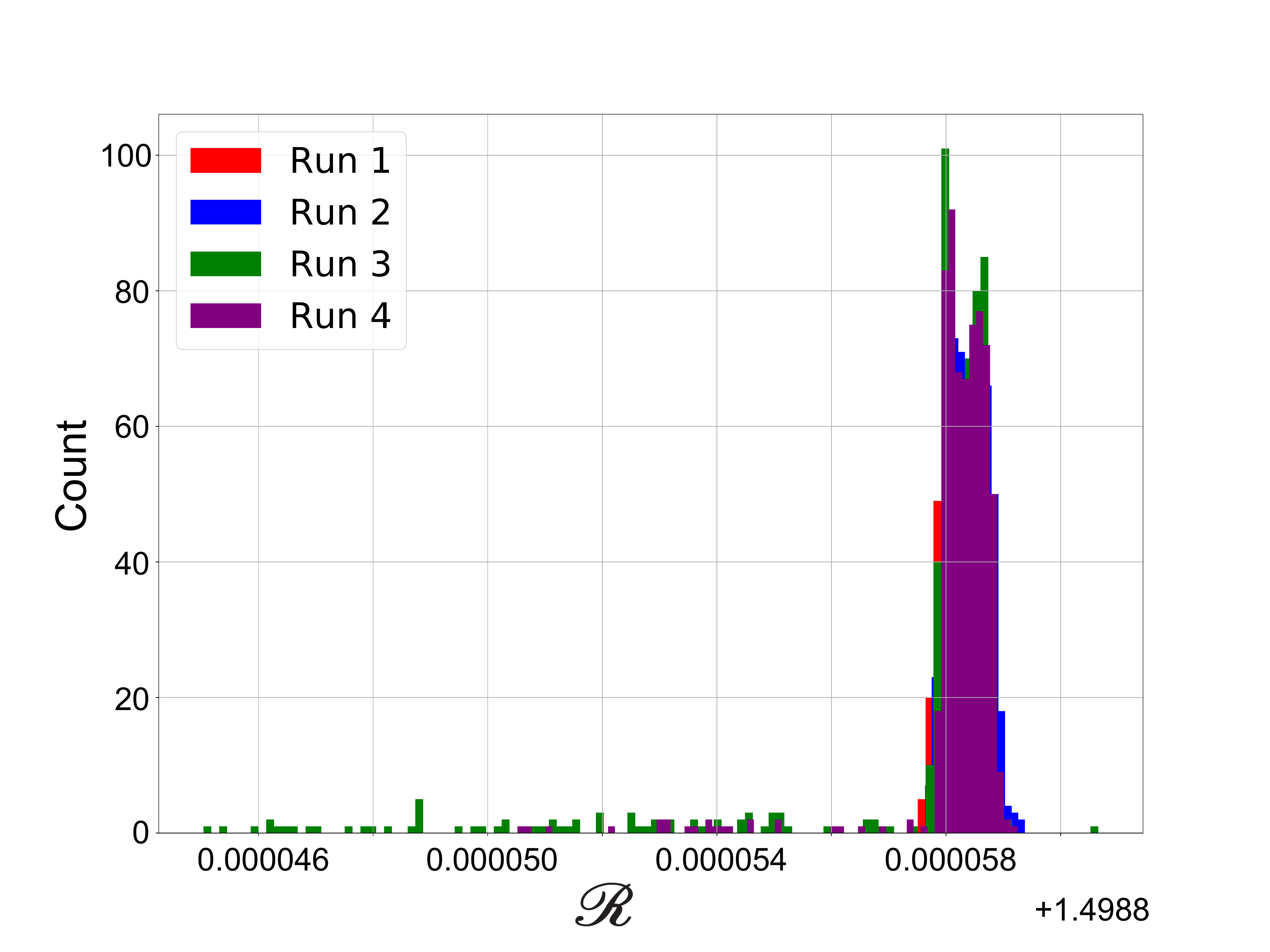}
\caption{The upper plot shows $\Omega_{87}$ versus $\Omega_{85}$ as determined by fits to the Fourier transform of the time-dependent optical rotation data for experimental Runs 1-4. The analysis procedure used to derive $\Omega_{87}$ and $\Omega_{85}$ is described in detail in Refs.~\cite{Kim17-GDM,Kim13-GDMsetup}. The data are taken with $\mb{B}$ along $+\mb{z}$ at the low-field magnitude of $B = 19.052729(3)~{\rm mG}$ using the alkene-coated cell. The lower plot shows a histogram of $\mc{R} = \Omega_{87}/\Omega_{85}$ for the positive field data from Runs 1-4.}
\label{Fig:scatter-histogram}
\end{figure}

Figure~\ref{Fig:scatter-histogram} shows the complete set of fit results for $\Omega_{87}$ and $\Omega_{85}$ from Runs 1-4 for $\mb{B}$ along $+\mb{z}$. The data shown in Fig.~\ref{Fig:scatter-histogram} include both that for which the pump beam was right-hand-circularly (RHC) polarized and left-hand-circularly (LHC) polarized. It is evident from inspection that Runs 3 and 4 have many more points exhibiting glitch behavior as compared to Runs 1 and 2, and note also that the glitches bias the data toward lower values of $\mc{R}$ as suggested by the data shown in Fig.~\ref{Fig:GradientCompensationExample}. The histograms of the data, shown in the lower plot of Fig.~\ref{Fig:scatter-histogram}, also exhibit non-Gaussian behavior due to shifts of $\mc{R}$ between data acquired using $LHC$ and $RHC$ pump light. These shifts are primarily due to the systematic effect related to scattered pump light discussed in the previous section and in Ref.~\cite{Kim17-GDM}. The net effect of the error due to scattered pump light is significantly reduced in the average of the data, with the residual unknown shift accounted for in the evaluation of systematic errors given in Table~\ref{Table:systematic-errors}.

The main idea of the approach to the data analysis in this work is that the data, in principle, can be partitioned into two groups: one set consistent with a linear relationship between $\Omega_{87}$ and $\Omega_{85}$ (denoted $S1$) and a second set exhibiting glitches that are statistically inconsistent with such a linear relationship (denoted $S2$). We assume in the analysis that the data from Runs 1 and 2 are all in $S1$, and then use a linear-regression-based technique from machine learning inspired by the perceptron model to divide the remaining data into the two distinct sets $S1$ and $S2$ (see, for example, the discussion of related methods in Ref.~\cite{witten2016data}). The assumption is that data belonging to $S1$, obeying a linear relationship between $\Omega_{87}$ and $\Omega_{85}$, are less likely to be biased by glitches. Furthermore, the data belonging to $S1$ should have larger values of $\mc{R}$ because the glitches systematically bias the data toward lower values of $\mc{R}$. Then the ratio $\mc{R}$ is calculated based only on the data from $S1$.

\begin{figure}
\includegraphics[width=3.5 in]{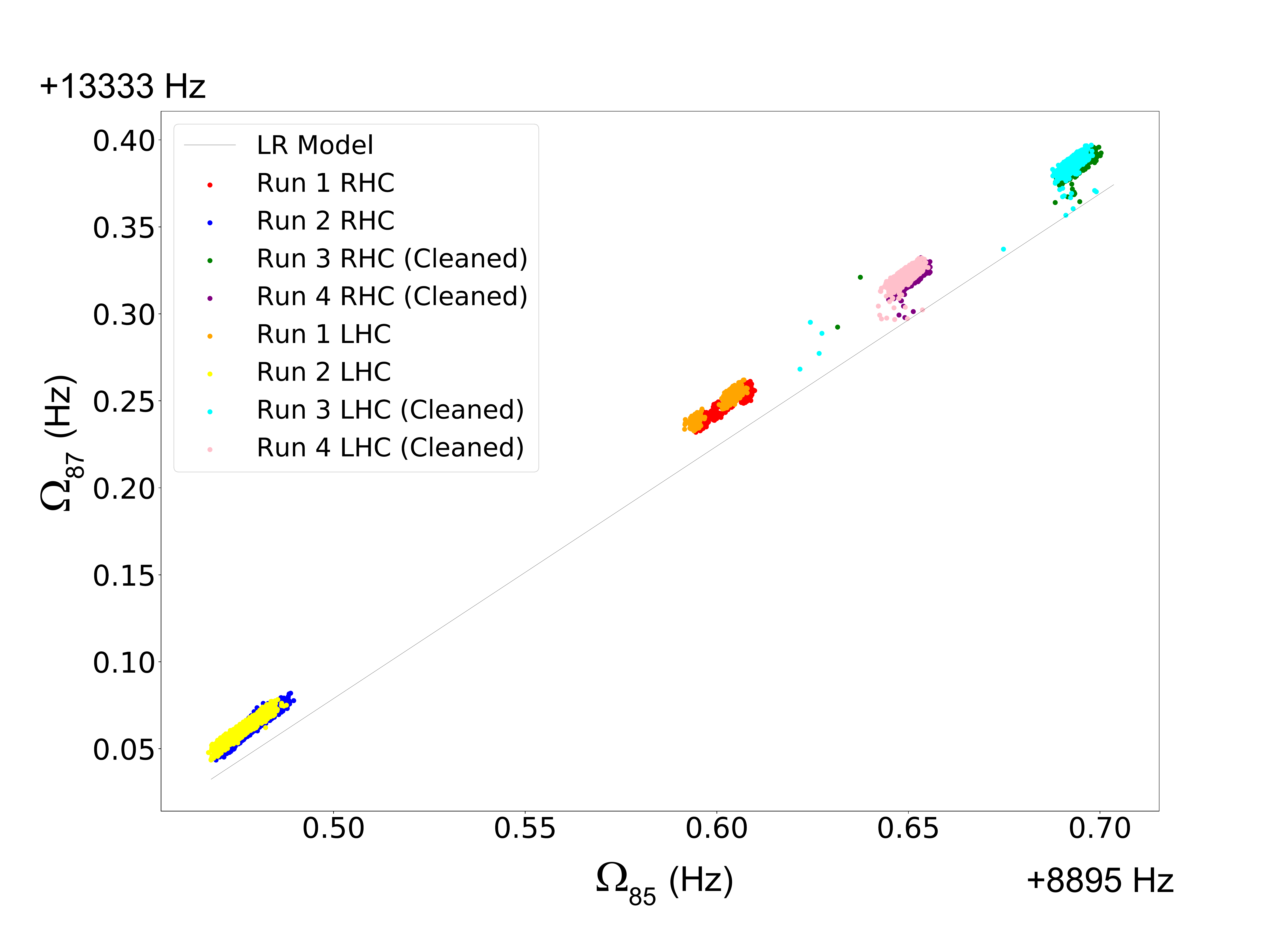}
\caption{Subset of data ($S1$, see text) used to calculate $\mc{R}$. The line labeled ``LR Model'' represents the line $\sigma$ below the line of best fit through the entire data set as determined by the machine-learning technique based on linear regression. Only data above the LR Model line (labeled ''cleaned'' for Runs 3 and 4) are used to used to calculate $\mc{R}$ (compare with Fig.~\ref{Fig:scatter-histogram}). }
\label{Fig:finalcut}
\end{figure}

To perform the linear regression, a subset of 70\% of the total data set is selected at random to be used as a training set. The training set is fit to a straight line, and the linear equation $f$ derived from the fit becomes the model. Then the remaining 30\% of the data is used to evaluate the accuracy of the model based on the mean-squared error. This process is repeated 1000 times to find the average line of best fit $f$ for the entire data set of $\Omega_{87}$ as a function of $\Omega_{85}$. The standard deviation $\sigma$ of the data set is also calculated from the fit residuals for Runs 3 and 4 as a measure of the distribution in the presence of glitches. Comparing data for which $\Omega_{87} \geq f$ to data for which $\Omega_{87} < f$, it is observed that the data for which $\Omega_{87} \geq f$ is consistent with a linear relationship between $\Omega_{87}$ and $\Omega_{85}$ with considerably smaller standard deviation as compared to the data for which $\Omega_{87} < f$, as could be expected by visual inspection of the data presented in the upper plot of Fig.~\ref{Fig:scatter-histogram}. As a conservative choice, in the end it was decided that all data for which $\Omega_{87} \geq f - \sigma$ would be partitioned into $S1$ and the rest of the data would constitute $S2$. In this manner, data with deviations $> \sigma$ below the line of best fit are discarded. The choice to partition the data set in this way was made so as to discard only data with statistically significant glitches to be more rather than less inclusive of suspect data. Thus the statistical errors of the measurements of $\mc{R}$ are more likely to be overestimated rather than underestimated. The mean and uncertainty in $\mc{R}$ are computed from the data in $S1$. Figure~\ref{Fig:finalcut} shows $S1$ for Runs 1-4. By inspection, it is seen that this procedure has removed a large fraction of the suspect data and preserved the bulk of the data that is consistent with the linear relationship between $\Omega_{87}$ and $\Omega_{85}$ as assumed. This same procedure was used on Runs 5-9.

\begin{figure}
\includegraphics[width=3.35 in]{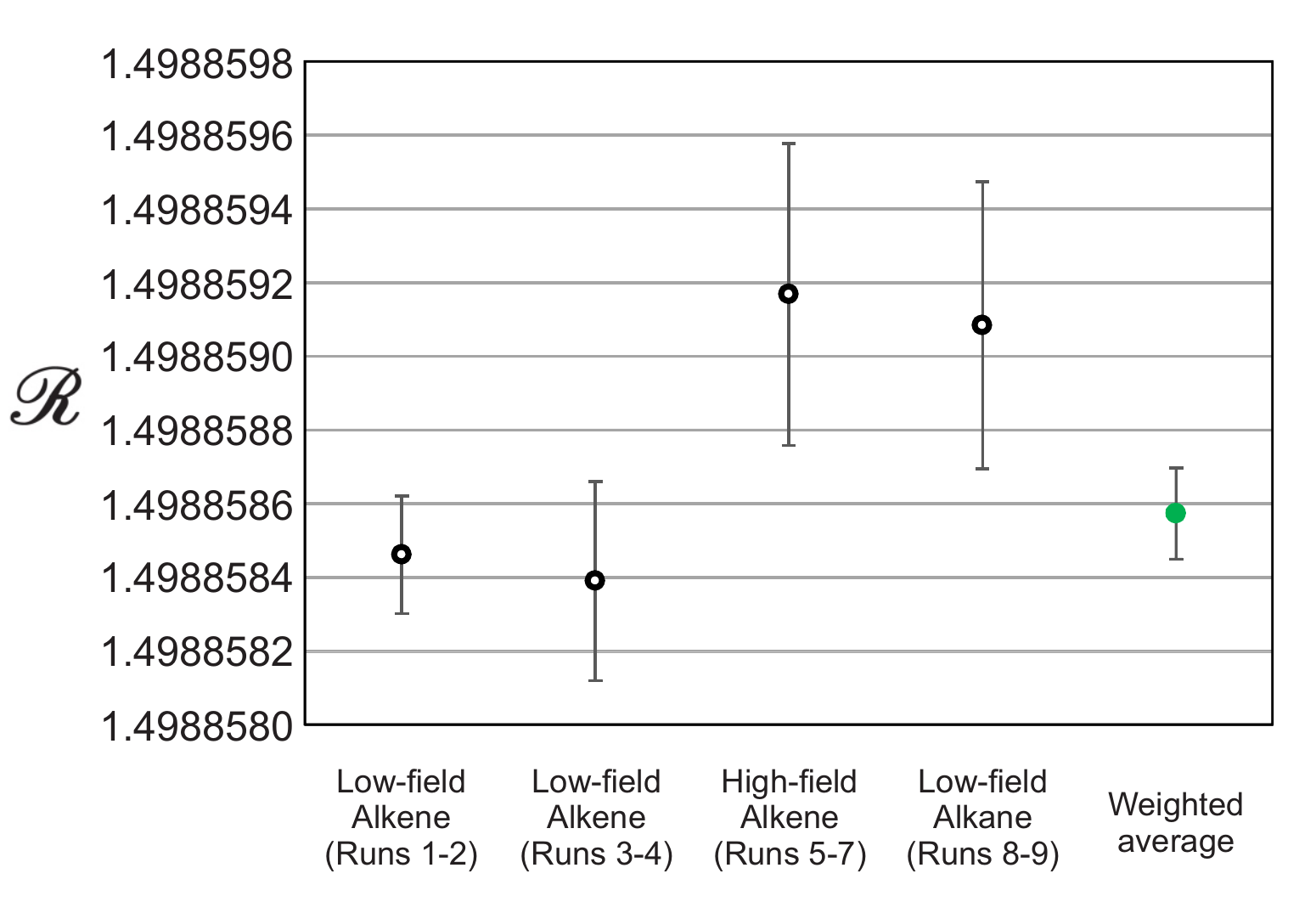}
\caption{Measurements of $\mc{R} = \Omega_{87}/\Omega_{85} = g_F(87)/g_F(85)$ for $^{87}$Rb atoms in the $F=2$ ground-state hyperfine level and $^{85}$Rb atoms in the $F=3$ ground-state hyperfine level, respectively.}
\label{Fig:combined-results}
\end{figure}

It is of interest to compare the partitioning of the data by the machine-learning technique used in this work to the partitioning of the data based on the fit uncertainty used in Ref.~\cite{Kim17-GDM}. A rough comparison of the two methods can be made based on the percentage of data belonging to $S1$, where again $S1$ is the data determined to have sufficiently small perturbations from glitches such that the data can be used in the computation of the final measurement result. Both methods assigned 100\% of the data from Runs 1 and 2 to $S1$. The fit uncertainty cut used in Ref.~\cite{Kim17-GDM} assigned 65.7\% of the positive field data and 45.8\% of the negative field data from Runs 3-9 to $S1$. The machine-learning technique used in this work assigned 94.4\% of the positive field data and 78.6\% of the negative field data from Runs 3-9 to $S1$. The general conclusion is that the partitioning method based on the fit uncertainty used in Ref.~\cite{Kim17-GDM} rejects significantly more data, which may be attributable to the fact that many of the effects that lead to poor fits do not actually affect $\sR$. Thus, at least in this case, the data sorting scheme based on a global analysis of the specific variables under investigation was more inclusive than the more generic sorting scheme based on quality-of-fit.

Figure~\ref{Fig:combined-results} shows the combined results from all measurements of $\mc{R}$. No statistically significant deviation of the processed data of Runs 3-9 from the data of Runs 1 and 2 is observed. Furthermore, the weighted average of all the data is consistent with the data from Runs 1 and 2. This suggests that after the processing of the data from Runs 3-9 using the technique from machine learning based on linear regression discussed above, the power-supply glitches do not significantly affect the determination of the ratio $\mc{R}$. Additionally, the different cells, cell coatings, cell orientations, and magnetic field magnitudes explored in Runs 1-9 do not exhibit statistically significant deviations from one another, supporting the view that these parameters do not cause any observable systematic errors.  The value of $\mc{R}$ from our measurements is found to be
\begin{align}
\mc{R} = \frac{g_F(87)}{g_F(85)} = 1.4988586(1).
\label{Eq:g-factor-ratio-measurement}
\end{align}
where the overall uncertainty is dominated by statistical error.

\begin{figure}
\includegraphics[width=3.35 in]{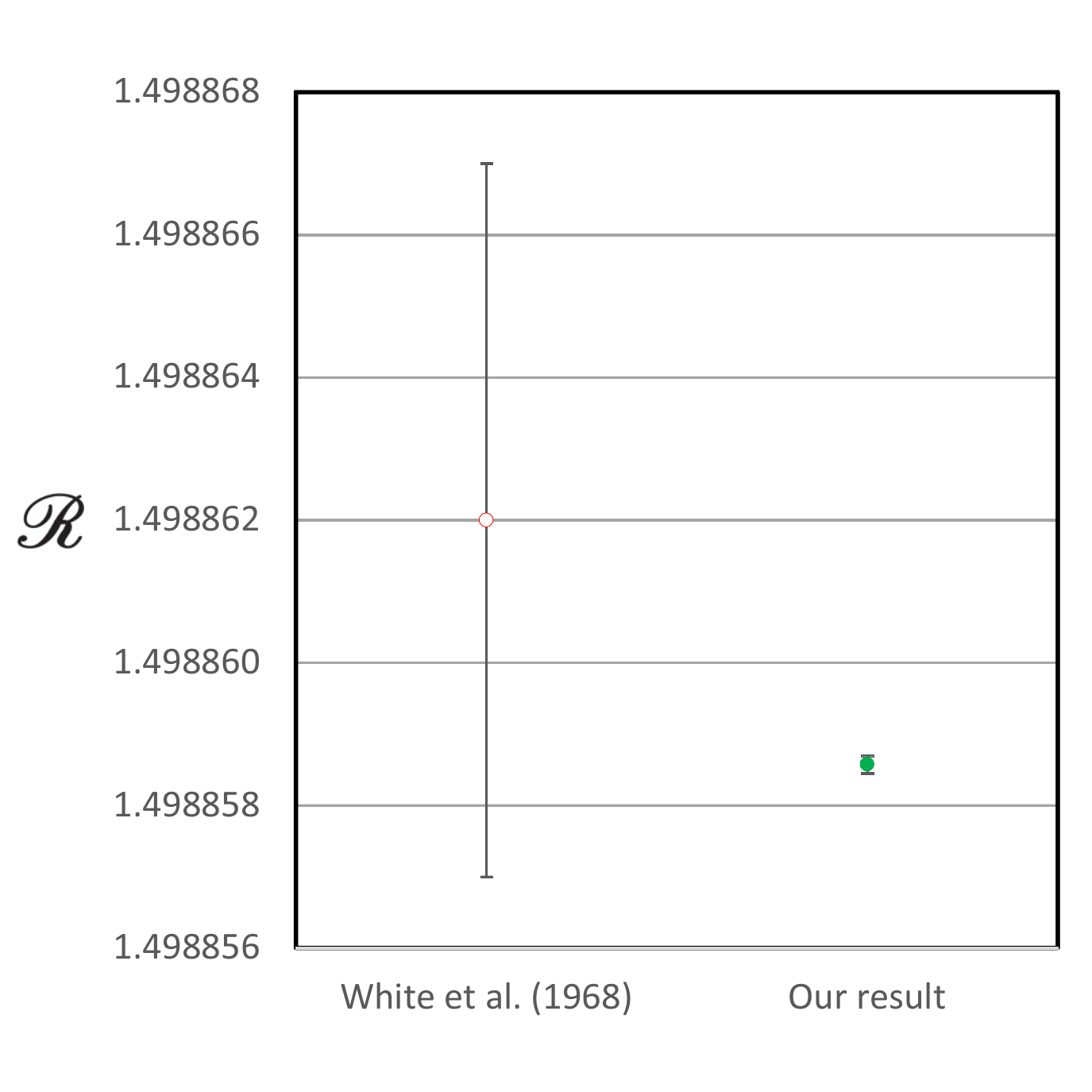}
\caption{Comparison of the results of the measurement of $\mc{R}=g_F(87)/g_F(85)$ in this work to the value reported in Ref.~\cite{Whi68-Rb-g-factor-ratio-measurement}.}
\label{Fig:results-comparison}
\end{figure}

Figure~\ref{Fig:results-comparison} compares the results of the measurements reported here to the value of $\mc{R}$ from Ref.~\cite{Whi68-Rb-g-factor-ratio-measurement}. It is seen that the two measurements are consistent with one another within experimental uncertainty. The uncertainty in our determination of $\mc{R}$ is $\approx 50$ times smaller than that of Ref.~\cite{Whi68-Rb-g-factor-ratio-measurement}.

\section{Conclusions}
\label{Sec:Conclusion}

In conclusion, the ratio between the Land\'e g-factors of the $^{87}$Rb $F=2$ and $^{85}$Rb $F=3$ ground state hyperfine levels was measured to be $g_F(87)/g_F(85) = 1.4988586(1)$, reducing the uncertainty in $g_F(87)/g_F(85)$ by a factor of $\approx 50$ compared to the previous best measurement of $g_F(87)/g_F(85) = 1.498862(5)$ reported in Ref.~\cite{Whi68-Rb-g-factor-ratio-measurement}. The measurement reported here, with an accuracy surpassing the 100 parts-per-billion level, is sensitive to effects related to quantum electrodynamics \cite{Ant94-Relativistic-corrections-to-g-factors,Chan11-cold-Rb-g-factor-expt}. This result may be useful as a test of atomic structure calculations \cite{karshenboim2005precision,roberts2015parity,nataraj2008intrinsic} and for accurate determination of magnetic fields \cite{Bud13}.

\begin{acknowledgements}
The authors are sincerely grateful to Kathryn Grimm and Dmitry Budker for useful discussions. We are also deeply indebted to generations of undergraduate students who spent countless hours over many years constructing and testing the experimental apparatus and acquiring the data used in this work, especially Rene Jacome, Ian Lacey, Jordan Dudley, Yan Li, Dilan Patel, Jerlyn Swiatlowski, Eric Bahr, Srikanth Guttikonda, Khoa Nguyen, Rodrigo Peregrina-Ramirez, Lok Fai Chan, Cesar Rios, Caitlin Montcrieffe, Claudio Sanchez, and Swecha Thulasi. We also thank Mohammad Ali, Li Wang, Valeriy Yashchuk, Valentin Dutertre, Mikhail Balabas for technical work on specific parts of the apparatus. This work was supported by the National Science Foundation under grant PHY-1707875.
\end{acknowledgements}

\bibliographystyle{andp2012}
\bibliography{Rb-g-factor-ratio}

\end{document}